\documentstyle[12pt]{article}
\newcommand{\pr}{\paragraph{}}
\newcommand{\be}{\begin{equation}}
\newcommand{\ee}{\end{equation}}
\newcommand{\bea}{\begin{eqnarray}}
\newcommand{\eea}{\end{eqnarray}}
\newcommand{\nd}[1]{/\hspace{-0.6em} #1}
\newcommand{\nk}{\noindent}
\begin{document}
\begin{titlepage}
\vspace{.1in}
\begin{flushright}
CERN-TH.6595/92\\
ACT-17/92 \\
CTP-TAMU-58/92 \\
\end{flushright}

\begin{centering}
\vspace{.1in}
{\Large {\bf String Theory Modifies Quantum
Mechanics }} \\
\vspace{.2in}
{\bf John Ellis}, {\bf N.E. Mavromatos} and {\bf D.V.
Nanopoulos}$^{\dagger}$   \\
\vspace{.05in}
Theory Division, CERN, CH-1211, Geneva 23, Switzerland  \\

\vspace{.05in}
\vspace{.05in}
\vspace{.1in}
{\bf Abstract} \\
{\small
\paragraph{}
We argue that the light particles in string theory obey an
effective quantum mechanics modified by the inclusion of a
quantum-gravitational friction term, induced by unavoidable couplings to
unobserved massive string states in the space-time foam. This term is
related to the $W$-symmetries that couple light particles to massive
solitonic string states in black hole backgrounds, and has a formal
similarity to simple models of environmental quantum friction. It
increases apparent entropy, and may induce the wave functions of
macroscopic systems to collapse.}
\end{centering}
{\small
}
\par
\vspace{0.4in}
\vspace{0.1in}

\par
\vspace{0.4in}

\begin{flushleft}
CERN-TH.6595/92 \\
ACT-17/92 \\
CTP-TAMU-58/92 \\
July 1992 \\
\end{flushleft}

\vspace{0.4in}

\noindent $^{\dagger}$ {\it Permanent address} :
Center for Theoretical Physics, Dept. of Physics,
Texas A \& M University, College Station, TX 77843-4242, USA,
and  \\
Astroparticle Physics Group,
Houston Advanced Research Center (HARC),
The Woodlands, TX 77381, USA.\\

\end{titlepage}
\newpage
\section{Introduction}
\pr
The discoveries of quantum mechanics and general relativity have
caused two of the greatest revolutions in twentieth-century physics,
and their reconciliation remains one of its most important pieces of
unfinished business. One could expect that this reconciliation would
entail a modification of one or both of these very successful
theories, and hope that it would cast light on the transition between
classical and quantum physics. Indeed, the only candidate for a
consistent quantum theory of gravity is string
theory \cite{gswbook}, which is
essentially non-local and hence modifies general relativity at short
distances. It has also been suggested that the usual formulation of
quantum mechanics and quantum field theory might require
modifications in a consistent quantum theory of gravity.
Specifically, studies of field theory in topologically non-trivial
space-times such as black hole backgrounds have indicated that
information loss across an event horizon requires the introduction of
mixed quantum-mechanical states, and the possibility, forbidden by
conventional quantum mechanics or $S$-matrix theory, that pure states
may evolve into mixed states.
\pr
It has been proposed by Hawking \cite{hawk} that a full quantum theory of
gravity might be formulated only in terms of density matrices
describing in general mixed states, and that the transitions between
initial and final density matrices might not be factorizable as
products of $S$-matrix elements and their hermitian conjugates:
\be
           \rho_{out} = \nd{S}
           \rho_{in} \qquad : \qquad  \nd{S}
           \ne   S^{\dagger} S
\label{I}
\ee

\nk
Evidence for this suggestion was
inferred from studies of
topologically non-trivial solutions of Einstein's equations in
Euclidean space-times. Two of us (J.E. and D.V.N.) then suggested
together with J.S. Hagelin and M. Srednicki \cite{ehns}
that the evolution of
quantum-mechanical systems over time-scales that are long compared
with the Planck time should be described by a modified Liouville
equation:
\be
      \partial _t \rho =i[\rho, H] + \nd{\delta H} \rho
\label{II}
\ee

\nk
A modification of the Liouville equation of the type (\ref{II})
is characteristic of open quantum-mechanical
systems \cite{open}, and represents in
our interpretation the intrinsic coupling of a microscopic system to
space-time foam.
\pr
We derived upper bounds of the order
of $1 GeV/ M_{Planck}$ on the magnitude of matrix elements of the
$\nd{\delta H }$ term in hadrons from the consistency with conventional
quantum mechanics of measurements of the $K^0 - {\overline K}^{0}$
system \cite{ehns,emnnextpaper}
and long-baseline
neutron interferometry \cite{ehns}. Subsequently, the
same two of us
together with S. Mohanty \cite{emhn}
demonstrated that the non-quantum-mechanical
effects of the $\nd{\delta H}$ would be enhanced in macroscopic
systems such as SQUIDs, and could lead to the collapse of the wave
function of a macroscopic object. An operationally similar
modification of the Liouville equation was
proposed independently
on completely
phenomenological grounds by Ghirardi, Rimini and Weber
\cite{ghirardi}, and the
required values of their model parameters were entirely consistent
with our upper bounds and order-of-magnitude estimates.
\pr
The study of topologically non-trivial space-times, event
horizons and singularities in string theory was opened up by
Witten's realization that suitable Wick rotations of a cosmological
string theory could be interpreted as black holes in Minkowski or
Euclidean space. We then embarked on a series of studies of quantum
coherence in such stringy black hole backgrounds, with the explicit
motivation of checking the conjectured breakdowns (\ref{I}) and
(\ref{II}) of conventional quantum field theory and quantum mechanics.
We found that quantum coherence was maintained in
the scattering of
light particles on a black hole background by an infinite set of
local $W_{1+\infty}$-symmetries \cite{emn1,emn2}, which
link light asymptotic
states to massive string states, and whose associated conserved
$W$-charges could encode all initial-state
information. We believe
that quantum coherence can
be maintained in this situation only in
an intrinsically
non-local theory such as string with its infinite set of gauge
symmetries, and that local field theories are doomed to failure in
this respect because they only have a finite set of hair. We have
shown that the infinite set of stringy $W$-hair is sufficient to label
all the black hole states, and thereby quench all the entropy
$S = M^2$ of a spherically-symmetric black hole in four
dimensions \cite{emn4}. We
have also argued \cite{emn3}
that black hole decay is a purely quantum-mechanical
higher-genus effect that does not require a thermal description in
terms of Hawking radiation, and that all the $W$-charges are in
principle measurable via scattering experiments or Aharonov-Bohm
phases for massive string
states \cite{emnw}. We have also given a stringy
desription of space-time foam in terms of a plasma of defects on the
world-sheet, which is intimately related to a Hall
conductor \cite{emndua}.
\pr
The maintenance of quantum coherence in the scattering of light
particles off a stringy black hole does $not$, however, mean that
conventional quantum mechanics and quantum field theory are
sacrosanct. The essential problem is that although all the $W$-charges
are {\it in  principle} measurable, this would entail observations
of the massive string states that are linked to the light states by
the $W$-symmetries, and such observations are not $in~ practice$
possible in realistic laboratory
experiments. We argue in this paper that a
modification of the Liouville equation of the type (\ref{II}) is
indeed $essential$ for observable systems
containing only light particles. Its
origin is exactly the $W$-linkage between the
light and heavy string states, which means that the former must be
regarded as an $open~ system$ coupled to the
unobserved heavy
degrees of
freedom. A modification of the form (\ref{II}) would be forbidden for
any {\it exactly  marginal} deformation of the underlying
string theory. However, it
is known from explicit examples that $in~ non-trivial ~space-time~
backgrounds$ such exactly marginal deformations involve massive string
states. Therefore, any deformation of the effective theory of light
particles in which the massive modes are unobserved will $necessarily$
not be exactly marginal, and will hence be associated with a
contribution to $\nd{\delta H}$ (\ref{II}). The unitarity of the
effective field theory of the light degrees of freedom and
Zamolodchikov's c-theorem \cite{zam}, as
proved by one of the authors (N.E.M.)
and Miramontes \cite{mavmir}, guarantee
that any contribution to $\nd{\delta H}$
can only
increase the entropy of the light-particle system. Topologically
non-trivial space-time backgrounds appear $intrinsically$ in the
stringy realization of space-time foam, so we
conclude that  a $\nd{\delta H }$ term
is $inevitable$ in string theory. We argue, moreover,
that the dominant such modifications are just those associated with
the known $S$-wave deformations of spherically-symmetric black holes.
A primitive order-of-magnitude estimate indicates that their
contribution to $\nd{\delta H }$ might be suppressed by just one
inverse power of $M_{Planck}$, and hence be
close to the present
experimental upper limits \cite{ehns,emnnextpaper}, and
in the ball-park needed to explain the
collapse of the wave function for a classical system.

\section{Renormalization Group Flow as Quantum Friction}
\pr
    Let us consider a general dynamical system whose state
is described by a density matrix $\rho\{q^i,p_i\}$, where the $q^i$
are generalized coordinates, and the $p_i$ are their associated
conjugate
momenta. In conventional quantum mechanics, the $q^i$
include conventional space-time coordinates, which in
string theory become parameters of background target spaces
corresponding to $\sigma$-models on the world-sheet. Thus the $q^i$
can be regarded as couplings in a space of possible $\sigma$-models.
We consider the evolution of $\rho$ with respect to a renormalization
group flow variable $t = \ln \Lambda$, where $\Lambda$ is some
covariant cutoff. We will identify $t$ with a conventional time
variable, which may or may not be considered as a string Liouville
mode. On dimensional grounds, the target time is then
measured in units of Planck length. This in turn
implies
that
even marginal changes in the renormalisation group scale
produce appreciable time-variations in target-space.
It is clear that renormalisability
of the system implies
\be
\frac{d\rho}{dt} = 0 = \frac{\partial  \rho}{\partial  q^i}
 \dot q^i   +
\frac{\partial  \rho}{\partial  p_i}\dot p_i
\label{III}
\ee

\nk where $ (.\dot ..)$ denotes $\partial _t (...)$.
\pr
It is known \cite{ems}
that in the Wilson renormalization scheme the
renormalization flow of the coordinates ($\sigma$-model couplings)
$q^i$ is given in the neighbourhood of a fixed point $S_0$ by
the gradient of Zamolodchikov's $c$-function $\Phi (q^i, S_0$):
\be
             \beta ^i \equiv \dot q^i
    =G_{ij}(S_0)\frac{\partial \Phi (q^i, S_0)}
{\partial q^i }
\label{IV}
\ee

\nk
where $G_{ij}$ is the metric in the space of coupling constants:
\be
          G_{ij} \equiv 2|z|^2 < O_i(z,{\overline z}) O_j(0, 0) >
\label{V}
\ee

\nk with $<...>$ denoting an
average with respect to the deformed
$\sigma$-model action $I^* + \int d^2z q^i O_i $, where
$I^*$ is the
fixed-point conformal field theory action,
and the $O_i$ are a complete set of (renormalised)
vertex operators.
\pr
    In view of equation (\ref{IV}), we can regard $\Phi$ as a Lagrange
function, to which we can associate a Hamiltonian in the usual way:
\be
         \Phi = \int dt
       [ \dot q^i\cdot p_i - H(q^i,p_i)  ]
\label{VI}
\ee

\nk
Again in the usual way, it follows that
\be
\dot q^i =\frac{\partial H}{\partial p_i}
\label{VIIa}
\ee

\nk and
\be
\dot p_i = -\frac{\partial  H}{\partial q^i} - G_{ij}\beta^j
\label{VIIb}
\ee

\nk The
second term in equation (8) is a friction term characteristic of
open dynamical systems \cite{open}, which
will be at the root of our subsequent
modification of quantum mechanics
\footnote{In the Wilson renormalisation scheme (\ref{IV}) the
friction is {\it linear }, which simplifies many of the
computations. However in practice, or in certain cases where
this scheme is not applicable \cite{ems}, one can work in schemes
where (\ref{IV}) is valid but
$G_{ij}=G_{ij}(q)$, in which case one is faced with a
{\it non-linear } friction problem. For most of our discussion
we can stay in the general case.}.
Substituting the expressions
(\ref{VIIa}),(\ref{VIIb})
into the time derivative equation (\ref{III}), we find the following
equation
\be
   \dot \rho = i[\rho, H] + G_{ij}\beta ^j
   \frac{\partial \rho}{\partial p_i}
\label{VIII}
\ee
for the classical evolution of the density matrix
\footnote{A similar
equation has been considered by Kogan \cite{kog}, but
his formalism is incorrect
because he
considers
total time derivatives in places
where one should consider partial ones.
In addition, he
identifies the Liouville field with the
target time coordinate, which clearly does not apply to
the two-dimensional black hole
string theory \cite{aben,polch}. We stress
again that from our point of view the world-sheet cutoff
is identified with the target time (evolution parameter)
independently of any (possible) identification of the
Liouville
field with time.}. When proposing any
such modification of the traditional evolution equation for the
density matrix, it is important to check that the total probability
$P = \int dp_i dq^i tr \rho ( p_i, q^i )$ is conserved. In our
case (\ref{VIII}), it is easy to check that
\be
      \partial _t P     = \int dp_l dq^l tr
\frac{\partial }{\partial p_i }(G_{ij}(q)\beta ^j (q) \rho (p,q))
\label{IX}
\ee

\nk which vanishes if there are no contributions from the boundary of
phase space, as would occur if it has no boundaries, or if $\rho$
vanishes there.
\pr
    The quantum version of this classical equation is obtained by
first replacing
\be
             \{,\} \rightarrow  \frac{1}{i} [,]
\label{X}
\ee
The quantum version of the second term in (\ref{VIII}) is obtained by
recalling Euler's equation (\ref{VIIa}), whose quantum version is
\be
        \frac{\partial H}{\partial p_i}
        = - i [q^i,H]
\label{XI}
\ee

\nk from which it follows that
\be
    \frac{\partial F(H)}{\partial  p_i} = -i [q^i,F(H)]
\label{XII}
\ee

\nk for any function $F$ of the Hamiltonian, including in particular the
density matrix $\rho$. We therefore arrive at the folowing evolution
equation for the quantum density matrix:
\be
  \dot \rho = i[\rho, H] -i G_{ij} [\rho , q^i]\beta ^j
\label{XIII}
\ee

\nk
We will not enter here into a discussion of the appropriate quantum
ordering of the factors in the second term in equation (\ref{XIII}),
which represents the quantum friction inherent to our open dynamical
system.
\pr
An additional remark we would like
to make concerns
the connection of the gradient flow relation (\ref{IV})
with the problem of a supersymmetric quantum mechanical
particle moving in a constant magnetic field ( more
specifically in a Kahler potential
$\Phi (q)$  \cite{vaf}). Indeed, it can be shown \cite{vaf}
that whenever
relation
(\ref{IV}) is valid, the evolution of $q^i$ in
$t=ln\Lambda$
can be described by a super-quantum mechanics
Hamiltonian  $H=\frac{1}{2}(Q + Q^{\dagger})$,
with supercharge $Q=\psi ^i (\nabla _i - \beta_i)$
and its hermitian
conjugate $Q^{\dagger}$, where the
$\psi ^i $ are
Grassmann
variables satisfying canonical commutation relations. The
charges are nilpotent : $Q^2=(Q^{\dagger})^{2}=0$ if
and only if
the {\it covariant} $\beta _i $ are curl-free, i.e. if relation
(\ref{IV}) is satisfied. In this case the metric in coupling
constant space can be viewed as a Kahler metric, whose
potential coincides with the flow-function $\Phi (q)$ \cite{vaf}.
This analogy
makes clearer the physical interpretation
of the renormalisation group flow as a
friction problem related
to the quantum
motion of
a particle in a magnetic field \cite{cald}.
\pr
    Our previous demonstration that the total probability $P$ is
conserved carries over directly from the classical case (\ref{VIII}) to
the quantum case (\ref{XIII}). However, entropy is $S = -
tr\rho\ln\rho$ is not necessarily conserved: we see easily that
\be
     \dot S     = - trGij[\rho,q^i]\beta^j ln\rho
\label{XIV}
\ee

\nk which does not in general vanish if $\beta^i \ne   0$. Our next task
is to demonstrate that indeed $\beta_i \equiv G_{ij}\beta ^j
\ne 0$, and in so doing we will
be able to argue that the entropy $S$ increases monotonically, as one
would expect on general physical grounds.

\section{Light Particle Operators are not Exactly Marginal}
\pr
    Since string theory is based on conformal field theory, one might
naively expect that the renormalization coefficients $\beta^i$
introduced above all vanish, and hence that the quantum friction
term in equation (\ref{XIII}) vanishes identically. This is indeed the
case in a flat space-time background, but is not true for operators
that create light particles in non-trivial backgrounds. The prototype
for this phenomenon is the operator creating a massless particle in a
two-dimensional string black hole background, which can also be
interpreted as a spherically-symmetric four-dimensional black hole
background. We first present a heuristic argument that $\beta ^i \ne
0$ in this case, which is followed by a more complete treatment
in section 5.
\pr
    We recall that the black hole solution possesses a
$W_{1+\infty}$-symmetry that is global on the world-sheet, and includes
the Virasoro algebra. It becomes a local symmetry
when elevated
in target space,
providing an infinite-dimensional extension of the general coordinate
transformations of general relativity. This $W$-symmetry is
responsible, in our interpretation, for the existence of the
$S$-matrix for scattering massless particles, confusingly
called ``tachyons'', in this black hole background, and hence
for the maintenance of quantum coherence in this scattering process.
The feature of these $W$-symmetries that is crucial for the present
discussion is that they link together states with different masses.
In particular, they relate massless ``tachyon'' states to massive
string levels. One could therefore suspect that the conformal
subalgebra of the $W$-symmetry associated with the Virasoro algebra
might also combine massless and massive states in an essential way, and
we shall see in a moment that such is indeed the case. Therefore,
the $\beta$-functions for operators contructed out of light
particle fields alone will not have the full $W$-symmetry, and
hence not be conformal in general, and hence have some $\beta ^i
\ne 0$. But physical laboratory experiments are conducted with light
particles, corresponding to states that are massless in the
approximation
which we discuss here, and do not measure any properties of the
massive string states. Therefore the relevant vertex operators are
precisely those that involve only massless states, and hence have
non-vanishing $\beta$-functions. Hence the $effective$ quantum
mechanics of observable particles has the $extra$ quantum friction term
in (\ref{XIII}), even though the density matrix of the $full$ string
theory would be described by the $usual$ Liouville equation without
such a term. The light observable particles constitute an open
quantum-mechanical system coupled by $W$-symmetry to unobserved
massive states.
\pr
    Examples of the above assertion that in non-trivial backgrounds
exactly marginal deformations generally involve massive string
states have been provided in ref. \cite{ChLyk}. In two-dimensional
flat-space Liouville theory, the following ``tachyon'' vertex operator is
exactly marginal:
\be
             \phi^{c,-c}_{-1/2,0,0} =(g_{++}g_{--})^{-\frac{1}{2}}
F(\frac{1}{2} ; \frac{1}{2} ; 1 ; \frac{g_{+-}g_{-+}}{g_{++}g_{--}})
\label{XV}
\ee

\nk where $g_{ab}$, $a,b =+,-$ represent the components
of a generic $SL(2,R)$ element \cite{witt}.
However, the corresponding exactly marginal operator in a
two-dimensional Minkowski space-time black hole, described by an
SL(2,R)/O(1,1) Wess-Zumino (WZ)
coset model, is
\be
          L^1_0 {\overline L_0}^1
          = \phi^{c,-c}_{-1/2,0,0}+ i (\psi ^{++}-
\psi^{--}) + \dots
\label{XVI}
\ee

\nk
where
\be
  \psi^{\pm\pm}
  \equiv : ({\overline J}^{\pm} )^N (J^{\pm})^N
  (g_{\pm\pm})^{j+m-N} :
\label{XVII}
\ee

\nk with
$J^{\pm} \equiv (k-2) (g_{\pm\mp}\partial _z g_{\pm\pm} -
g_{\pm\pm}\partial _z g_{\pm\mp} )$, and
${\overline J}^{\pm} \equiv (k-2) (g_{\mp\pm}\partial _{{\bar z}}
g_{\pm\pm} - g_{\pm\pm}\partial _{{\bar z}} g_{\mp\pm})$,
with $k$ the WZ model level parameter \cite{witt}.
The combination $\psi^{++} - \psi^{--}$ generates a level-one
massive string mode, and the dots in equation (\ref{XVI})
represent
operators that generate higher-level massive string states. An
analogous exactly marginal opearator is
\be
          L^2_0 {\overline L_0}^2   = \psi^{++} + \psi^{--} + \psi^{-+}
+\psi^{+-} + \dots
\label{XVIII}
\ee

\nk
which also involves in an essential way operators for massive string
modes. For later use we note that the coupling constant,
$\alpha $,
corresponding to this deformation of the coset model
is responsible for a global rescaling of the target
space-metric \cite{ChLyk}, and therefore to a global
constant shift by $\alpha $ of the dilaton field. Thus it
produces shifts in the black hole mass \cite{witt}.
\pr
We now remark that in two-dimensional string theory
all these massive
states are in fact discrete states of the type
generated by the $W_{1+\infty}$ algebra which maintains quantum
coherence \cite{emn1,ms}. Each
of these discrete solitonic states can be represented
as a singular gauge configuration \cite{klepol},
whose conserved $W$-charges can be
measured in principle by generalized Aharonov-Bohm phase
effects \cite{emnw}.
However, laboratory measurements of light microscopic objects such as
the $K^0 - {\overline K}^0$ system or interfering neutron beams do not
measure such effects, so do not observe the massive states and are
restricted to the light (massless) parts of the exactly marginal
operators (\ref{XVI},\ref{XVIII}). These by themselves are not exactly
marginal, hence the corresponding light-field $\beta$-functions do not
vanish, and therefore make non-zero contributions to the quantum
friction term in equation (\ref{XIII}). We shall give order
of maginitude estimates of these effects in the next section.
\pr
    As we have discussed in ref.\cite{emndua}
the sum over
quantum configurations in the string path integral includes a sum
over many vortices and spikes on the world-sheet:
\be
Z=\int D{\tilde X} exp(-\beta S_{eff}({\tilde X}) )
\label{act}
\ee

\nk where ${\tilde X} \equiv   \beta^{\frac{1}{2}}X$,
$\beta $ is a `temperature ' variable  characterising
the topological defects on the world-sheet,
and
\bea
\nonumber
\beta S_{eff} &=& \int d^2 z [ 2\partial {\tilde X}
{\overline \partial } {\tilde X} +  \frac{1}{4\pi }
[ \gamma _v\epsilon ^{\frac{\alpha}{2}-2}
(2 \sqrt{|g(z)|})^{1-\frac{\alpha}{4}}: cos (\sqrt{2\pi \alpha }
[{\tilde X}(z) + {\tilde X}({\bar z})]):   \\
&+&  (\gamma _v, \alpha,
{\tilde X}(z) + {\tilde X}({\bar z}) )
\rightarrow (
\gamma _m, \alpha ', {\tilde X}(z) - {\tilde X}({\bar z}))]]
\label{XIX}
\eea

\nk Here $\gamma_{v,m}$ are the fugacities for vortices and spikes
respectively, and
\be
 \alpha \equiv 2\pi \beta q_v^2  \qquad \alpha ' \equiv
\frac{e^2}{2\pi \beta}
\label{anom}
\ee

\nk are related to the conformal
dimensions $\Delta_{v,m}$ of the vortex and
spike creation operators respectively, namely
\bea
\nonumber
\alpha =4 \Delta _v \qquad   \alpha ' =4 \Delta _m \\
      \Delta _m =\frac{(eq_v)^2 }{16 \Delta _v}
\label{conf}
\eea

\nk In the
low-temperature phase relevant in the present-day Universe,
this path-integral is dominated by a plasma of world-sheet
spikes corresponding
to microscopic Minkowski-space black holes, which constitute a
stringy realization of space-time foam \cite{emndua}.
Therefore, many
configurations with non-zero values of the $\beta^i$ contribute
importantly to the string path integral, and one cannot $a~ priori$
expect the modification (\ref{XIII}) of the quantum Liouville equation
to be negligible.
\pr
The reader might worry that although individual configurations
contribute to the quantum fiction term in (\ref{XIII}), some as yet
unknown symmetry principle might cause their total net contribution to
vanish. An immediate intuitive counterargument to this suggestion
is the comment that no-one has ever seen a macroscopic body speed
up as a result of friction!
Thus a cancellation would be surprising in the light of our
physical picture of
the renormalisation
group flow (of unitary theories)
as
the motion of a (supersymmetric) particle in
coupling constant space, under the influence of an external
magnetic field \cite{vaf,ems}.
Indeed, as we shall now explain, the
unitarity of the truncated effective light-particle field theory
guarantees that all quantum friction terms tend to increase the
entropy, in accord with everyday experience, and hence cannot cancel in
the path-integral sum.
\pr
In our case,
it is easy to see that
the renormalization group flow is
irreversible, as has also been argued in ref. \cite{emndua}
on the basis of
the isomorphism of the space-time foam theory of vortices and spikes
with a Hall conductor. The rate of change of the Lagrange function
$\Phi$ (6) is given by
\be
     \dot \Phi     =  \beta^i \frac{\partial \Phi }
     {\partial q^i}
\label{XX}
\ee

\nk which can be rewritten using equation (\ref{IV})
\be
    \dot \Phi     = -  \beta^i  G_{ij} \beta^j
\label{XXI}
\ee

\nk Since the low-energy effective field theory of the light degrees of
freedom should be unitary by itself, the metric $G^{ij}$ in the space
of coupling constants must be positive definite \cite{zam}.
Hence the Lagrange
function decreases monotonically, corresponding to a monotonic
change in the effective central charge. This in turn corresponds to
a monotonic increase in the entropy, as can be seen explicitly from
the expression (\ref{XIV}) for
the rate of change of the quantum entropy.
The commutator factor in (\ref{XIII}) can be rewritten as
\be
        -i[ \rho ( H ), q^i ] =
\frac{\partial \rho}{\partial p_i} =
\frac{\partial \rho}{\partial H} \dot q^i
=\frac{\partial \rho}{\partial H} \beta^i
\label{XXII}
\ee

\nk Substituting this expression into equation (\ref{XIV}), we find
\be
 \dot S = Tr \beta^i G_{ij} \beta ^j
\frac{\partial \rho}{\partial H} ln\rho
\label{XXIII}
\ee

\nk In systems that exchange energy with their environment,
as is our case, the density matrix is actually given by
\be
      \rho =   e^{ \beta(F-H)}
\label{fren}
\ee

\nk where $F$ is the free energy of the system. In the
particular case of strings, $F$ is identified
with the effective action $\Phi $ \cite{mavmir}.
Thus, taking into account the facts
that $ln \rho < 0$,
and that for unitary $\sigma$-models
$\beta ^i G_{ij} \beta ^j = 2|z|^4 <\Theta (z,{\bar z}),
\Theta (0)>$ is positive definite \cite{zam}, one observes that
the quantum entropy does indeed increase monotonically
thanks to the unitarity of the effective low-energy theory. This
simple and general argument excludes the possibility of a sneaky
cancellation between different contributions to the quantum friction.
\pr
    We conclude this section by noting an essential difference
between wormhole calculus and the type of quantum gravitational
physics that we are discussing in this paper. Wormholes connect
different parts of the target space-time via throats. Classical
gauge symmetries that are carried by particles falling into one end
of the wormhole are carried out by particles expelled at the other end,
and there is no sign of information loss. Wormholes can give rise to
non-local effects in space-time, which may be physically dubious for
other reasons, but do not cause obvious problems for conventional
quantum field theory or quantum mechanics. We do not address here
the question whether
string theory admits such wormholes. The defects on the world-sheet
that we discuss here correspond to Minkowski space black holes without
throats, where, as discussed above,
information is transferred monotonically by
$W$-symmetries to unobserved massive string modes, resulting in the
above-mentioned monotonic increase in the apparent entropy of the
observable light degrees of freedom.

\section{Simple Models of Quantum Friction Effects}
\pr
    Before making an order-of-magnitude estimate of quantum friction
effects in the low-energy effective field theory derived from string,
we first remind the reader of the standard formalism for microscopic
systems interacting with ``environmental'' oscillator modes. This
enables us to make contact with the previous literature on the
transition between the quantum behaviour of microscopic systems and
the classical behaviour of macroscopic systems. The standard
``environmental'' formalism assumes weak perturbations around the
equilibrium state of the microscopic system, which is justified in
our case by the suppression of Planck-mass excitations at low
energies. The effective lagrangian for a single harmonic oscillator
coupled to ``environmental'' oscillators representing the infinite set
of massive string modes is
\be
    L=\frac{1}{2}M\dot q^2 -V(q) + \frac{1}{2} \sum_j
(m_j \dot x_j^2 - m_j \omega _j^2 x_j^2 ) -
\sum_j F_j(q)x_j - \sum_j \frac{F_j(q)^2}{2m_j \omega_j^2}
\label{tn}
\ee

\nk where the last two terms represent the interactions. The density
matrix of the system (\ref{tn}) can be expressed as a path integral in
the usual way. The reduced density matrix describing the evolution
of the
primary microscopic harmonic oscillator
from $q=q_i$ at time $t=0$ to $q=q_f$ at time $T$
is then given by
\be
K(q_i,q_f, T  ) =K_0(T) \int _{q(0)=q_i}^{q( T  )=q_f}
Dq(\tau) e^{-S_{eff}[q(\tau)]/\hbar }
\label{trt}
\ee

\nk where the prefactor
$K_0(T)$
involves only the environmental oscillator frequencies \cite{cald},
and can be absorbed in the normalisation of the
uncoupled case.
The Euclidean
effective action $S_{eff}$
is given by
\be
  S_{eff}[q(\tau)] =\int _0^{T} d\tau
  (\frac{1}{2}M\dot q^2 + V(q))
+ \frac{1}{2} \int _{-\infty}^{+\infty} d\tau ' \int _0^{T}  d\tau
\alpha (\tau - \tau ') (q(\tau ) - q(\tau ') )^2
\label{trtI}
\ee

\nk where
\be
   \alpha (\tau -\tau ') \equiv \sum _n
   \frac{C_n^2}{4m_n \omega _n} e^{-\omega _n (\tau -\tau ')}
\label{trtII}
\ee

\nk In deriving these results, we have simplified matters by assuming
linearity in the interactions: $F_j(q) = q C_j$,
corresponding to a
Wilson renormalisation scheme
in string theory \cite{ems}. In this case it can be shown that
$\alpha$ is given by the asymptotic form of the so-called
Drude model \cite{cald}
\be
    \alpha (\tau - \tau ') = \eta \frac{1}{(\tau -\tau ')^2}
\label{drude}
\ee

\nk where $\eta$  is the ``environmental'' friction
coefficient, given by
\be
      \eta   \equiv \sum_j \frac{C_j^2}{m_j\omega _j^2 } \delta (\omega
- \omega _j)
\label{frcc}
\ee

\nk We see from Eq. (\ref{trtII}) that the strengths of the dissipation
terms are suppressed for heavy ``environmental'' oscillators, a feature
that we might expect to carry over to more massive modes in string
theory. The ``environmental'' dissipation effects (\ref{trtII}) are
suppressed linearly in the large masses or frequencies.
If
this result is
carried over
to string theory (with the correspondence of the oscillator
frequencies $\omega _j$
in (\ref{frcc}) to the string mass levels),
the
magnitude of the stringy quantum friction
effects could be comparable to the upper bounds established in
refs. \cite{ehns,emnnextpaper} for
the neutral kaon and neutron systems.

\section{Order-of-Magnitude Estimate in String Theory}
\pr
In
previous sections we have seen that in two-dimensional
string theory the exactly marginal operators that
turn on backgrounds for the light degrees of freedom
in the presence of a black-hole
contain necessarily massive string states which
are discrete solitonic
states, that are not measured in laboratory
experiments. To estimate the orders of magnitude of the quantum-
gravitational friction term in (\ref{XIII}) and
of the rate of change of the
entropy (\ref{XXIII}), we need
to discuss the magnitude of the renormalization
functions $\beta^i$. In general, one has
\be
   \beta ^i = \sum _{N=1}^{\infty} C^i_{i_1...i_N}g^{i_1}...g^{i_N}
\label{beta}
\ee

\nk where the
coefficients $C^{i}_{i_1...i_N}$
are not
totally symmetric among covariant and contravariant
indices. The usual on-shell string N-point amplitudes
coincide with the expansion coefficients
$\gamma _{i_1...i_N}=<\int d^2z_1V_{i_1}...\int d^2z_NV_{i_N}>$
of the
covariant $\beta _i \equiv G_{ij}\beta ^j$ \cite{mavmir2,brust}.
These coefficients
are totally symmetric
as follows from the flow equation (\ref{IV}).
Factorization
implies that all higher-order coefficients are obtainable from 3-point
functions \cite{brust}. If
one chooses an appropriate Wilson renormalization scheme,
all contact terms are eliminated from these 3-point functions if the
boundaries of moduli space are treated correctly \cite{distdoy}, as
we assume for the
coset black hole model, in which case the metric tensor $G^ij$ in
coupling constant space
is flat with its non-zero entries $O (1)$ \footnote{
The existence of such a Wilson scheme, and hence the
vanishing of such contact terms, is subtle in certain
cases
such as supersymmetric
strings \cite{green}, requiring a
careful definition
of correlation functions with colliding
punctures
\cite{distdoy}. Thus it might not be easy
for this scheme
to be implemented
in practice
in the black hole case,
whose singularity is associated with a
twisted supersymmetric fixed point \cite{eguchi}.
If a conventional renormalisation scheme
were adopted in this case \cite{verl}, it would
give non-linear friction. However,
we believe that physical observables
would not be affected, since the corresponding
conformal field theory is well-defined
even at the singularity \cite{witt}.}.
\pr
For exactly marginal operators with both light- and
heavy-state terms: $V_i  =  V_i^{(L)}    +  V_i^{(H)} $, such as
$L^1_0 {\overline L}^1_0$ (17), we have
\be
  0 =\beta^i = G^{ij}(S_0)\sum_{N=1}^{\infty}
   g^{i_1}...g^{i_N} <\Pi_{n=1}^{N}(V_{i_n}^{(L)}+ V_{i_n}^{(H)})>
\label{thrf}
\ee

\nk Therefore, defining the light-state renormalization coefficients
\be
     {\hat \beta}^i = \sum_{N=1}^{\infty}
      G^{ij}(S_0) g^{i_1}...g^{i_N}
<\Pi_{n=1}^{N}
 V_{i_n}^{(L)}>
\label{thrfiv}
\ee

\nk we find
\be
   {\hat \beta}^i = - \sum _{N=1}^{\infty}
    \sum_{M=1}^{N}
   g^{i_1}...g^{i_N} <\Pi_{m=1}^{M}
    V_{i_m}^{(M)}\Pi_{n=1}^{N-M} V_{i_n}^{(L)}>
\label{thrsix}
\ee

\nk The effective $\beta^i$ appearing in sections 2 and 3 are in fact the
${\hat \beta}^i$ (\ref{thrfiv}) in
the light-field theory, and we now revert to the hatless
notation of those earlier sections. As we have seen, these
effective $\beta ^i$
are non-vanishing to
the extent that the mixed light-heavy N-point functions in (36) are
non-zero.
\pr
To estimate these, we use the Euclidean counterpart
of the $SL(2,R)/O(1,1)$  WZ coset model \cite{witt}, which can
be mapped \cite{BerKut} into a
$c =1$ matrix model with a modified cosmological constant on the
world-sheet \footnote{Although Euclidean and Minkowski coset models
are not physically equivalent in this formalism
\cite{emn3,distl},
the Euclidean version
will be sufficient
for our
order-of-magnitude estimates if
we assume that the
analytic
continuation in target space needed to produce
Minkowski black holes
does not affect
the scaling arguments we are using below.}: in their notation
\be
            2\pi\mu \beta {\overline \beta }
 exp (-2\frac{\phi}{\alpha_{+}}) \equiv  \mu  V_{1,1}^{(-)}
\label{twentyone}
\ee

\nk where $\beta$,${\overline \beta }$ are chiral bosons
used along with the free field $\phi $ and another
chiral boson $\gamma $
to parametrise $SL(2,R)$ \cite{BerKut}.
The scale $\mu$ is related to the mass of the black hole by
$\mu  =  M_{BH} / M_{Planck}$, and the massive string states are
represented in general by
\be
      V_{r_1r_2}^{(\pm)} =[\partial ^{r_1r_2}X + ...]
e^{i\frac{r_1 - r_2}{\sqrt{2}} \pm \beta _{r_1r_2}^{\pm}\phi }
\label{twentifive}
\ee

\nk The $(-)$ states are often discarded in usual Liouville
theory \cite{polch}, but
play a key role here, as we shall see. In particular, following
ref. \cite{BerKut}, we have the following
representation for the WZ model representing a two-dimensional
black hole \cite{witt}
\be
     S_{WZ} = S_{c=1,\mu=0} + \mu V_{1,1}^{(-)}
\label{twentysix}
\ee

\nk The (+) states are massive string modes, so the correlation functions
of interest to us can be expressed as
\be
  \sum_{l=0}^{+\infty} \frac{1}{l!}\mu^l
  <\Pi _{i=1}^{N'}V^{(+)}_{r_ir_{i+1}}
\Pi _{j=1}^{M} V_j^{0} \Pi _{m=1}^{l} V_{r_{m-1}r_{m}}^{(-)}>_1
\label{twentyseven}
\ee

\nk where
the superscript (0)
denotes a massless ``tachyon'' state, and the
subscript 1 on the v.e.v. denotes the fact that it is calculated in
a $c$=1 matrix model with $\mu  = 0 $.
\pr
    We now seek to understand which correlation
functions are in general non-vanishing, and how they scale
with $\mu$. The first important observation in this regard is that
correlators of $V^{(-)}$
operators alone vanish: only correlators with
at least one $V^{(+)}$ operator are non-zero. This means that the black
hole dynamics necessarily turns on physical massive string states, as
we had seen previously from the point of view of the $W$-symmetry
linking massless and massive modes \cite{emn1,emnw}. The
second comment is that the
non-zero correlators contain in general logarithmic factors that
violate the expected power-law scaling with $\mu$. Specifically,
there are factors of $(ln \mu )^{\pm 1}$
for insertions of $V^{(\pm)}$
operators respectively, associated with divergences close to the
Fermi level and representing physical pole conditions in amplitudes
\cite{grosnewm,difkut}. Thus we
have the following general scaling laws:
\be
<\Pi _{i=1}^{2N'}V^{(+)}_{r_{i-1}r_{i}}
\Pi _{j=1}^{M} V_j^{0} \Pi _{m=1}^{2N} V_{r_{m-1}r_{m}}^{(-)}>_1
\propto \bigg \{ \begin{array}{cc}
\mu^s (ln\mu )^{N'-N} \times F(\{ k \}) ~~for N'-N > 0 \\
0 \qquad otherwise        \end{array}
\label{twentyeight}
\ee

\nk where $F(\{ k \} )$ is a kinematical factor, and
the power $s$ is determined by naive kinematical scaling
considerations. It is related to
the Liouville-energy
non-conservation which arises
from the reality of the coupling
of the
Liouville field in the expression for the
vertex operators \cite{pol}. For
resonant amplitudes, $s=0$. For
general $s$
one can only compute the
amplitudes by analytic continuation from the
integer $s$ case
\cite{goul}. Such a procedure is assumed in the following.
\pr
    We see in the general scaling law
(\ref{twentyeight})
logarithmic factors in the non-zero correlators that
appear to
vanish as $ln \mu \rightarrow  0$, i.e.,
as $\mu \rightarrow  1$. Therefore we are led to
conclude that the
contributions to the quantum-gravitational
friction are suppressed for
configurations with $\mu \simeq 1 $. Formally, the
expression (\ref{twentyeight})
is ill-defined for $\mu  \rightarrow 0$, i.e.,
$ln \mu \rightarrow -\infty $. This
we interpret as a reflection of the well-known
ill-definition of the $c$ = 1 matrix model
itself when $\mu  = 0$. Analytic continuation is not
safe in this regime, since
if
one
formally evaluates the amplitudes
in the limit $\mu \rightarrow 0$ by analytically
continuing to $s=integer$ $>$ $N'-N$ in (\ref{twentyeight}), then
the amplitudes vanish. We interpret these subtleties
as follows:
as
discussed in ref. \cite{emndua},
we believe that the appropriate
string vacuum is a space-time foam consisting of a plasma of spikes on
the world-sheet corresponding to microscopic Minkowski-space black
holes. The end-state of black hole decay is a microscopic black hole
that is indistinguishable from this foam, so the latter should
be
subtracted and serve as a regulator for black hole physics. This
would therefore be dominated in our application by configurations with
$\mu  \ge 1 $ in the gravitational path integral
analogue of the simple model (32).
\pr
  Incorporating the estimate (\ref{twentyeight})
into
the expression (\ref{thrsix}) for the
effective light-field renormalization coefficients $\beta^i$, and then
using the path integral (20,21) to perform the sum over microscopic
Minkowski black holes in the space-time foam, which is dominated by
configurations with $\mu  =  M_{BH} / M_{Planck} \ge 1 $
as discussed
above, we estimate that the coefficients of the quantum-gravitational
friction terms in (\ref{XIII},\ref{XXIII})
are $O (1)$ in Planck units. Thus their effects
are suppressed by dimensional powers of $M_{Planck}$, and the largest
could be $O ( M_{Planck}^{-1})$ as in
the simple model (32). Their detailed
evaluation requires more knowledge of the effective low-energy theory
derived from string, and goes beyond the scope of this paper.
\newpage
\section{ Collapse of the wave-function}.
\pr
It has been argued \cite{emhn}
that the microscopic entropy increase
offered by quantum gravity
effects may cause the collapse
of macroscopic wave functions and their
transition from quantum to classical behaviour.
We now discuss whether quantum quantum-gravitational
friction (\ref{XIII}) and the rate of entropy increase
(\ref{XXIII})
could have this effect. We start from the simple
model for friction described in section 4. In the
string case, the renormalized $\sigma$-model
couplings play the r\^oles of coordinates
$q$, and $\tau$ is the renormalization group scale
parameter, identified with target time.
Since the $q$'s are almost but not exactly marginal
in the truncated effective light-mode field theory,
the dominant contributions to the path-integral
in the asymptotic Drude model (\ref{drude})
come from the limit
where $\tau \rightarrow \tau '$.
In this case eq (\ref{trtI}) may be written as
\bea
\nonumber     \rho (q_i,q_f,  T  ) / \rho_S (q_i,q_f,  T  ) \simeq
e^{ - \eta  \int _0^{T} d\tau \int_{\tau-\epsilon}^{\tau + \epsilon}
d\tau '
\frac{(q(\tau) - q(\tau '))^2}{(\tau - \tau ')^2}} \simeq \\
e^{-\eta \int_0^{T} d\tau \int_{\tau ' \simeq \tau }
d\tau '
 \beta ^i G_{ij}(S_0) \beta ^j } \simeq
e^{ - DT ({\bf q_i}  - {\bf  q_f} )^2 + \dots }
\label{asym}
\eea

\nk where $D$ is a small constant,
proportional to the sum of the
squares of the
effective anomalous dimensions of the renormalised
couplings $q^i$.
We
reinterpret
(31) as representing the overlap, after an elapsed time interval
$T$, between quantum systems localized at
different values ${\bf q}={\bf q_i}, {\bf q_f}$ of
the coordinates, and
the subscript ``S'' denotes quantities evaluated
in
Schroedinger
quantum mechanics.
Equation (\ref{asym})
exhibits a quadratic dependence on the $\beta $-functions
in this simple mechanical model for friction.  It
captures
all the essential features of the complicated
string case, in which the light-mode effective $\beta$-functions
are suppressed by inverse powers of $M_{Planck}$, as we
discussed in the previous section.
\pr
The vanishing of off-diagonal terms in the density matrix
leads in general to the collapse of the wave-function.
This effect is usually negligible for microscopic
light-mode systems - except possibly for special cases
such as the $K^{0}-{\overline K}^{0}$ system discussed
elsewhere \cite{ehns,emnnextpaper} - but it is enhanced
for macroscopic systems. This was first derived in ref.
\cite{emhn} in the context of a wormhole model, but the result
is more general and applies here. We consider a macroscopic
body containing many particles with coordinates $q^i, i=1,2,...N$.
Decomposing these into a centre-of-mass coordinate $Q$ and relative
coordinates $\Delta q^i$, $(i=1,2,...N)$, writing
the full density matrix as $\rho _q = \rho _Q \rho _{\Delta q}$,
and assuming friction terms $-Dt(q^i -q'^i)^2 $ $(i=1,2,...N)$,
the equation for the centre-of-mass motion is
\be
 \dot{\rho}_Q \simeq i[\rho _Q , H(Q)]\rho _{\Delta q} +
i [\rho _{\Delta q}, H(\Delta q)]\rho _Q - \sum_{i=1}^{N}
D(q^i - q'^i)^2\rho (q^i,q'^i)
\label{centre}
\ee

\nk Tracing over the relative coordinates $\Delta q^i$, we find
\be
\dot{\rho}_Q \simeq i[\rho _Q, H(Q)] - ND(Q-Q')^2\rho _Q
\label{macr}
\ee

\nk and hence
\be
    \rho (Q',Q,t) / \rho _S(Q',Q, t) \simeq
e^{-NDt(Q' - Q)^2 }
\label{final}
\ee

\nk where we see an enhanced suppression for bodies
containing many particles.
As an illustration, if $D \simeq m^6_{proton}/M^3_{Planck}$
as in ref. \cite{emhn}, then the locations of bodies with
$N \ge 10^{24}$ constituent particles (cf Avogadro's
number)
are fixed to within
a Bohr radius within about $10^{-7}$ sec. A similar mechanism
for wave function collapse was introduced phenomenologically
in ref. \cite{ghirardi} without reference to any microscopic
model.
\pr
A final comment we would like to make concerns other
possible
effects of the space-time foam, which could also
be enhanced
macroscopically. As we have argued in \cite{emndua}, the space-time
foam is represented as a plasma of world-sheet spikes, and one
should examine the motion of a single light string mode in this
plasma which
is not yet
feasible.
We notice, however, that taking into account the no-net-force
condition among the spikes of the plasma
\cite{emndua}, we can define a `centre-of-mass' coordinate
for the black hole moduli and examine the motion of the
tachyon in a single coset model, representing in some sense
this `collective' effect. In such a case, following \cite{emhn},
one would have an enhanced
macroscopic non-quantum mechanical
effect if there  were
microscopic violations of the no-net-force
condition, leading to a motion of the centre of mass of the
black-hole moduli. Such effects are left for future investigations.
However one should keep in mind that they might also
lead to ``observable'' modifications of quantum mechanics.

\section{Discussion}
\pr
    The following remarks may help the reader to understand
intuitively the relation of this work to other studies. As we have
emphasized previously \cite{emnw}, the
scattering of light particles off a black hole
resembles the Callan-Rubakov \cite{cr}
description of the scattering of charged
particles off a grand unified monopole. There, conservation
laws enforce selection rules on the scattered particles, and
here similar selection rules ensure that quantum coherence is conserved
by the full light/heavy particle system. The situation we have
discussed in this paper is more akin to the 't Hooft
\cite{thooft} analysis of
instanton effects in the Standard Model. In that case, scattering light
particles have a very small probability of ``meeting'' an instanton,
but then it generates new physics. In our case, light particles again
have a very small probability of ``meeting'' a microscopic or virtual
black hole, but then it also generates new physics. Since the heaavy
degrees of freedom are necessarily ignored in a light-particle
scattering experiment, information and hence coherence are lost in
this case.
\pr
    We have laid out in this paper the conceptual basis for the
modification of laboratory quantum mechanics in string theory. We
re-emphasize that the full string theory is perfectly
quantum-mechanical, but that experiments cannot see this unless they
measure all the massive string states, which are linked to light
states by an infinite set of $W$-symmetries. Observing these massive
modes is an impossibility in realistic laboratory experiments.
Therefore, one must work with an effective theory of the light degrees
of freedom as an open system, with the corresponding modification
(\ref{XIII})
of the Liouville equation for the density matrix. A simple extension
of this argument will
lead to a non-factorizing $\nd{S}$-matrix description
of the transition between initial and final density matrices in
quantum field theory. The quantum-gravitational friction that we have
identified in this paper appears able to explain qualitatively the
transition from quantum-mechanical behaviour of microscopic systems to
the classical behaviour of macroscopic systems containing Avogadro's
number of elementary particles.
\pr
\noindent {\Large {\bf Acknowledgements }} \\

The work of D.V.N.
is partially supported by DOE grant DE-FG05-91-ER-40633 and
by a grant from Conoco Inc.


\begin{thebibliography}{99}
\bibitem{gswbook} M.B. Green, J.H. Schwarz and E. Witten,
{\it Superstrings } Vol. I \& II (Cambridge University Press
1988).
\bibitem{hawk} S. Hawking, Comm. Math. Phys. 43 (1975), 199;
{\it ibid} 87 (1982), 395.
\bibitem{ehns} J. Ellis, J.S. Hagelin, D.V. Nanopoulos and
M. Srednicki, Nucl. Phys. B241 (1984), 381.
\bibitem{open} Y.R. Shen, Phys. Rev. 155 (1967), 921;
\par A.S. Davydov and A. A. Serikov, Phys. Stat. Sol.
B51 (1972), 57;
\par B.Ya. Zel'dovich, A.M. Perelomov, and V.S. Popov,
Sov. Phys. JETP 28 (1969), 308;
\par For a recent review see : V. Gorini {\it et al.},
Rep. Math. Phys. Vol. 13 (1978), 149.
\bibitem{emnnextpaper} J. Ellis, N.E. Mavromatos and D.V. Nanopoulos,
CERN and Texas A \& M Univ. preprint, CERN-TH.6596/92;
CTP-TAMU-47/92; ACT-12/92 (following paper).
\bibitem{emhn} J. Ellis, S. Mohanty and D.V. Nanopoulos,
Phys. Lett. B221 (1989), 113; {\it ibid} B235 (1990), 305.
\bibitem{ghirardi} G.C. Ghirardi, A. Rimini and T. Weber,
Phys. Rev. D34 (1986), 470; G.C. Ghirardi, O. Nicrosini,
A. Rimini and T. Weber, Nuov. Cim. 102B (1988), 383.
\bibitem{emn1} J. Ellis, N.E. Mavromatos and D.V. Nanopoulos,
Phys. Lett. B267 (1991), 465.
\bibitem{emn2} J. Ellis, N.E. Mavromatos and D.V. Nanopoulos,
Phys. Lett. B272 (1991), 261.
\bibitem{emn4} J. Ellis, N.E. Mavromatos and D.V. Nanopoulos,
Phys. Lett. B278 (1992), 246.
\bibitem{emn3} J. Ellis, N.E. Mavromatos and D.V. Nanopoulos,
Phys. Lett. B276 (1992), 56.
\bibitem{emnw} J. Ellis, N.E. Mavromatos and D.V. Nanopoulos,
Phys. Lett. B284 (1992), 27, 43.
\bibitem{emndua} J. Ellis, N.E. Mavromatos and D.V. Nanopoulos,
CERN and Texas A \& M Univ. preprint, CERN-TH.6534/92;
CTP-TAMU-47/92; ACT-12/92; CERN-TH.6536/92;
CTP-TAMU-48/92; ACT-13/92;
\bibitem{zam} A.B. Zamolodchikov, JETP Lett. 43 (1986), 730;
Sov. J. Nucl. Phys. 46 (1987), 1090.
\bibitem{mavmir} N. Mavromatos and J.L. Miramontes,
Phys. Lett. B212 (1988), 33; N. Mavromatos, Mod. Phys.
Lett. A3 (1988), 1079; Phys. Rev. D39 (1989), 1659;
A. Tseytlin, Phys. Lett. B194 (1987), 63; H. Osborn,
Phys. Lett. B214 (1988), 555.
\bibitem{ems} N.E. Mavromatos, J.L. Miramontes and
J.M. Sanchez de Santos, Phys. Rev. D40 (1989), 535.
\bibitem{kog} I. Kogan, Univ. of British Columbia preprint
UBCTP 91-13 (1991).
\bibitem{aben} I. Antoniadis, C. Bachas, J. Ellis and
D.V. Nanopoulos, Phys. Lett. B211 (1988), 393; Nucl. Phys.
B328 (1989), 117; Phys. Lett. (1991), 278.
\bibitem{polch} J. Polchinski, Nucl. Phys. B324 (1989), 123.
\bibitem{vaf} C. Vafa, Phys. Lett. B212 (1988), 29;
S. Das, G. Mandal and S. Wadia, Mod. Phys. Lett. A4 (1989), 745.
\bibitem{cald} R.P. Feynman and F.L. Vernon Jr., Ann. Phys.
(NY) 94 (1963), 118;
\par A.O. Caldeira and A.J. Leggett, Ann. Phys.
149 (1983), 374.
\bibitem{ChLyk} S. Chaudhuri and J. Lykken, FERMILAB preprint
FERMI-PUB-92/169-T.
\bibitem{witt} E. Witten, Phys. Rev. D44 (1991), 314.
\bibitem{ms} G. Moore and N. Seiberg, Int. J. Mod. Phys. A7
(1992), 2601.
\bibitem{klepol} I. Klebanov and A.M. Polyakov, Mod. Phys. Lett.
A6 (1991), 3273;
\par A.M. Polyakov, Princeton Univ. preprint PUPT-1289 (1991).
\bibitem{mavmir2} N.E. Mavromatos and J.L. Miramontes,
Phys. Lett. B226 (1989), 291.
\bibitem{brust} R. Brustein, D. Nemeschansky and S. Yankielowicz,
Nucl. Phys. B301 (1988), 224.
\bibitem{distdoy} J. Distler and M. Doyle, Princeton Univ.
preprint PUPT-1312 (1992).
\bibitem{green} M.B. Green and N. Seiberg, Nucl. Phys.
B299 (1988), 559.
\bibitem{eguchi} T. Eguchi, Mod. Phys. Lett. A7 (1992), 85.
\bibitem{verl} R. Dijkraaf, H. Verlinde and E. Verlinde,
Nucl. Phys. B371 (1992), 269.
\bibitem{BerKut} M. Bershadsky and D. Kutasov, Phys. Lett.
B266 (1991), 345.
\bibitem{distl} J. Distler and P. Nelson, Nucl. Phys. B374
(1992), 123.
\bibitem{grosnewm} D. Gross, I. Klebanov and M.J. Newman,
Nucl. Phys. B350 (1991), 621.
\bibitem{difkut} P. di Francesco and D. Kutasov, Phys. Lett.
B261 (1991), 385; Nucl. Phys. B375 (1992), 119.
\bibitem{pol} A.M. Polyakov, Mod. Phys. Lett. A6 (1991), 635.
\bibitem{goul} M. Goulian and M. Li, Phys. Rev. Lett. 66
(1991), 2051.
\bibitem{cr} C. Callan, Phys. Rev. D25 (1982), 2141;
\par V.A. Rubakov, Nucl. Phys. B203 (1982), 311;
\par V.A. Rubakov and M.S. Serebryakov, Nucl. Phys.
B237 (1984),239.
\bibitem{thooft} G. 't Hooft, Phys. Rev. Lett. 37 (1976), 8;
Phys. Rev. D14 (1976), 3432; {\it ibid} D18 (1978), 2199.
\end{thebibliography}
\end{document}